\documentstyle[prd,aps,12pt]{revtex}

\begin{document}

\draft
\title{Extraction of the Ratio of the N$^*$(1535)
Electromagnetic Helicity Amplitudes from Eta Photoproduction
off Neutrons and Protons}
\author{ Nimai C. Mukhopadhyay$^{a}$, J. -F. Zhang$^a$
and M. Benmerrouche$^{b}$\\
$^a$ Physics Department\\ Rensselaer Polytechnic Institute\\
Troy, NY 12180-3590, U.S.A. \\
$^{b}$ Linear Accelerator Laboratory \\ University of Saskatchewan\\
Saskatoon, SK S7N 5C6, Canada}

\maketitle

\begin{abstract}
Using the recent precise measurements of eta photoproduction in proton
and deuteron targets, we extract the ratio of the helicity amplitudes
$A_{1/2}^{n}/A_{1/2}^{p}$, for the excitation of
N$^*$(1535), in the effective Lagrangian approach.
It is fairly model-independent, free from the final-state
interaction effects, and negative as predicted by the quark
models. We stress the importance of polarization observables in
further elucidation of the N$^*$(1520) photoexcitation amplitudes.
\end{abstract}
\pacs{PACS numbers: 13.60.Le, 12.40.Aa, 25.10.$+$s, 25.20.Lj}

\newpage

The study of eta photoproduction near threshold off nucleons has
been established as a precision tool to study electrostrong
properties of the N$^*$(1535) resonance predominantly excited in
this process\cite{ref1,ref2}. This should allow us to
test an important prediction of the quark model\cite{ref3}, a
lack of flavor symmetry in the
amplitudes for exciting N$^*$(1535) off proton and neutron targets by
electromagnetic probes. Thus, the calculated ratio of the transverse
helicity amplitudes in the two targets is\cite{ref3}:
\begin{equation}
A_{1/2}^{n} = c A_{1/2}^{p},
\end{equation}
where the coefficient $c$ is predicted\cite{ref3} to be negative,
 with its
magnitude varing from about 0.7 to unity. Considerable relativistic
corrections are indicated\cite{ref3} in estimating these amplitudes,
but there is no theoretical dispute about the sign of $c$. This flavor
asymmetry of Eq.(1) has also important implications in photo- and
electroproduction of eta mesons in complex nuclei.

Main purpose of this Letter is to test the inequality in the magnitudes
of the above helicity amplitudes in the proton and neutron targets,
by appealing to the new, precise data of near-threshold photoproduction
of etas from the proton and deuteron targets at Mainz\cite{ref2,ref4}. We
use the proton data\cite{ref2} to fix   unknown
parameters of the strong sector of
our tree-level effective Lagrangian\cite{ref1},
and utilize the extracted neutron cross-sections\cite{ref4} to determine
the unknown $A_{1/2}$ helicity amplitude to excite the N$^*$(1535)
state, fixing the N$^*$(1520) electromagnetic helicity amplitudes by
the proton data and the quark model.
A comparison of this amplitude with that for proton would then yield the
test of the flavor dependence of the helicity amplitudes,
predicted by the quark model. We then retain
all strong interaction parameters and fit the rest of the parameters in
our effective Lagrangian approach (ELA) to
the neutron data. Finally,  we investigate the importance of
polarization observables in further sorting out of the N$^*$(1520)
photoexcitation amplitudes.

An inference of neutron cross-sections from the deuteron data\cite{ref4}
is far from being straightforward. The loosely bound neutron in
deuteron {\it is}
the most convenient target for eta photoproduction for this
purpose\cite{ref4}, {\it but nuclear corrections must be
made} to extract neutron cross-sections from those of deuteron\cite{ref5}.
Krusche {\it et al.}\cite{ref4} have obtained the best agreement
between their proton\cite{ref2} and deuteron\cite{ref4} data by
constraining the ratio of the neutron to proton cross-sections as
\begin{equation}
\sigma_{n}^{exp}/\sigma_{p}^{exp}=0.66\pm 0.07,
\end{equation}
for lab photon energies from eta production threshold up
to 792 MeV. This holds reasonably well both
for the differential and the total cross-section.  We shall use this
{\it experimentally inferred number} as a constraint for our effective
Lagrangian approach.

Our analysis reported here is different from that of Krusche
{\it et al.}\cite{ref4}. We take into account the possible
model-dependence in the contributions other than that for N$^*$(1535).
We extract the ratio of helicity amplitudes for N$^*$(1535) for
proton and neutron targets and show that this extraction can
be done in a model-independent fashion.

At the first sight, one might conclude that the experimental result
(2) straightforwardly implies (1), but this is not necessarily so.
Only in the naive approximation that the excitation
of N$^*$(1535) is the {\it sole} contribution to the eta photoproduction,
one can readily infer this connection. This is because
the non-resonant nucleon and vector meson exchange
contributions\cite{ref1}
play important but significantly
different roles in proton and neutron targets,
as do the excitations of other resonances
such as N$^*$(1520)\cite{ref1}. Investigations of these contributions
to the photoproduction of eta mesons off neutrons constitute an
important part of this work. Here we would defer from other
recent theoretical treatments of the process\cite{ref5,ref6}.
Many of these works also stress the unified treatment of pion and eta
channels. While this is theoretically desirable, current
uncertainties\cite{ref7}  of
the strong interaction ($\pi$, $\eta$) data base do not allow us
to carry out
such an ambitious treatment satisfactorily at present. Thus, we shall
 not
utilize the latter here.

We shall now discuss the theoretical ingredients of our ELA
to describe the photoproduction of eta off neutrons,
\begin{equation}
\gamma +n\rightarrow n+\eta ,
\end{equation}
near threshold. This discussion parallels our ELA treatment\cite{ref1}
of the process off protons, $\gamma +p\rightarrow p+\eta$. Thus, at the
tree-level, we can write the invariant matrix element $i{\cal M}_{fi}$
in the standard form\cite{ref1}
\begin{equation}
i{\cal M}_{fi}=\bar{u}_{f}(p_{f})\sum^{4}_{j=1}A_{j}(s,t,u)M_{j}u_{
i}(p_{i}),
\end{equation}
with $M_{1}=-1/2\gamma_{5}\gamma_{\mu}\gamma_{\nu}F^{\mu\nu}$,
$M_{2}=2\gamma_{5}P_{\mu}(q_{\nu}-k_{\nu}/2)F^{\mu\nu}$,
$M_{3}=-\gamma_{5}\gamma_{\mu}q_{\nu}F^{\mu\nu}$, $M_{4}=-2
\gamma_{5}\gamma_{\mu}P_{\nu}F^{\mu\nu}-2MM_{1}$, where
$p_{i}$, $p_{f}$ are the nucleon four-momenta, $k$ and $q$ are
the photon and the meson four-momenta, $P^{\mu}=(p_{i}+p_{f}
)^{\mu}/2$, $F^{\mu\nu}$ is the electromagnetic field tensor.
The {\it neutron} Born terms will give, for the meson-
nucleon pseudoscalar interaction\cite{ref8},
\begin{equation}
A^{ps}_{1}=A_{2}^{ps}=0,
\end{equation}
\begin{equation}
A_{3}^{ps}=-eg_{\eta}\frac{k_{n}}{2M}\left[\frac{1}{s-M^2}-
\frac{1}{u-M^2}\right],
\end{equation}
\begin{equation}
A^{ps}_{4}=-eg_{\eta}\frac{k_{n}}{2M}\left[\frac{1}{s-M^2}+
\frac{1}{u-M^2}
\right],
\end{equation}
with $k_{n}\simeq -1.91 nm$, $g_{\eta}$, the peudoscalar
$\eta NN$ coupling strength,
$M$, the nucleon mass. The above results follow from the
corresponding proton contributions, by using isospin symmetry
and noting the trivial differences between
 neutron and proton in charge and
anomalous magnetic moment. The corresponding results
 for the pseudovector
interaction is\cite{ref8}
\begin{eqnarray}
A_{1}^{pv}=eg_{\eta}\frac{k_
{n}}{2M^2}, & A^{pv}_{k}=A^{ps}_{k},
& k=2, 3, 4.
\end{eqnarray}
Here we utilize the pseudoscalar eta-nucleon coupling.

The t-channel vector meson exchange contributions for the neutron eta
photoproduction can be likewise obtained from the proton
case by the isospin symmetry. Given the fact that there is only
isovector contribution for the $\rho$ case, but not for the
$\omega$-exchange, the A coefficients for the $\rho$ exchange
amplitude has to be multiplied by $-1$, relative to the proton:
\begin{equation}
A_{i}^{\rho}(n)=-A_{i}^{\rho}(p),
\end{equation}
\begin{equation}
A_{i}^{\omega}(n)=A_{i}^{\omega}(p).
\end{equation}

For the resonance exchanges, the isospin symmetry of the
effective Lagrangian also allows us to construct the neutron amplitudes
from those of the proton\cite{ref1}. Thus, the proton coupling strengths
in terms
of transition amplitudes are $k^p_R =k^s_R +k^v_R$, while
those for the neutrons are:
\begin{equation}
k_{R}^{n}=k_{R}^{s}-k_{R}^{v},
\end{equation}
where $s$ and $v$ represent the isoscalar and isovector resonance
coupling strengths respectively. This completes
the transcription of the proton
couplings into the neutron ones. The complexity of
 the spin-3/2 resonance
propagators\cite{ref9} will be treated in the same way as
in the proton case\cite{ref1}, with the
parameters controlling the spin-1/2 sector unchanged from the
proton case, except when we need to fit the neutron data.
We stress the importance of this physics in our theoretical
considerations.

The general treatment of the neutron photoproduction requires,
for $E_{\gamma}$ varying from the eta threshold
($E_{\gamma}^{n}(th)=706.94
MeV$) to 1200MeV, the consideration of resonances N$^*$(1440),
N$^*$(1535), N$^*$(1520), N$^*$(1650) and N$^*$(1710).
However, the Mainz {\it proton data} can be adequately treated in an
ELA which contains only the nucleon Born terms, $\rho$, $\omega$
vector meson exchanges in the t-channel and the contributions of
N$^*$(1535) and N$^*$(1520), as shown by our nice fit to the Mainz
proton angular distribution data sampled in Fig.1, yielding $g_{\eta}$,
the pseudoscalar eta-nucleon coupling constant, to be
approximately $2.3$.
The contribution of
N$^*$(1535) {\it dominates}, but the roles of the other
contributions are {\it essential} in order to reproduce the observed
angular distributions\cite{ref2}. We shall use this effective version of
our ELA to {\it predict} the angular distributions for the eta
photoproduction from the neutron target. For this we take the value of $c$
in (1) to be equal to $-0.83$,
as suggested by recent quark model estimates, for example, by
Capstick\cite{ref3}. We also use the quark model as a guide to convert the
proton $A_{1/2}$, $A_{3/2}$ helicity amplitudes for the N$^*$(1520)
excitation off neutrons.

We  infer neutron data from deuteron observations\cite{ref4} as follows:
\begin{equation}
\frac{(d\sigma /d\Omega)_{n}}{(d\sigma /d\Omega )_{p}}\simeq
\frac{\sigma_{n}}{\sigma_{p}}\simeq\frac{2}{3}.
\end{equation}
In Table I, we give the E$_{0+}$ amplitudes for the proton and the
neutron targets for photoproduction of etas at threshold, obtained
from typical fits to the proton\cite{ref2} and our inferred neutron data.
We also compare our estimates with those of Tiator {\it et al.}\cite{ref6}.
Various non-resonant and resonant contributions can now be
compared for the
two targets. The dominance of the N$^*$(1535)
excitation emerges in all fits. It is, however,
a result of {\it different} model-dependent background
contributions in the two
targets.

We now display our predictions of typical angular distributions of the
eta photoproductions off neutrons in Fig. 1, along with our fits of the
Mainz proton data. The role of the resonances alone is also shown.
The predicted angular distribution for neutron
nicely matches the two-third ratio
between neutron and proton, inferred experimentally by
Krusche  {\it et al.}\cite{ref4} from their deuteron target
experiment. This suggests that {\it the value of $c$ equal to $-0.83$ is
consistent with the results of the Mainz deuteron experiment.}

We can now turn the last argument around. By demanding
a fit to the empirical observation (2) of Krusche
{\it et al.}\cite{ref4}, we can extract a parameter $\xi_{n}$ for
N$^*$(1535), defined\cite{ref1} in the usual notation, similar to
our extraction\cite{ref1} of $\xi_{p}$,
\begin{equation}
\xi_{n}=\sqrt{\chi_{n}\Gamma_{\eta}}A^{n}_{1/2}/\Gamma_{T},
\end{equation}
where the kinematic factor $\chi_{n}$ can be computed from the
neutron mass and other kinematic parameters of the N$^*$(1535)
excitation. We can characterize the electrostrong property of
N$^*$(1535), as inferred from the neutron experiment, to be
\begin{equation}
\xi_{n}= (-1.86 \pm 0.20)\times 10^{-4} MeV^{-1}.
\end{equation}
The error here includes an  estimate of our ELA model uncertainties.
{}From our analysis\cite{ref1} of the proton data\cite{ref2}, we have
\begin{equation}
\xi_{p}=(2.20  \pm 0.15)\times 10^{-4}  MeV^{-1}.
\end{equation}

We can now combine our inferences from the Mainz
proton and the deuteron experiments by taking the ratio
of the extracted parameters $\xi_{n}$ and $\xi_{p}$:
\begin{equation}
\frac{\xi_{n}}{\xi_{p}}=\sqrt{\frac{\chi_{n}}{\chi_{p}}}
\frac{A^{n}_{1/2}}{A^{p}_{1/2}}.
\end{equation}
{\it In the above ratio, the strong interaction property of the
N$^*$(1535),
arising from the decay of this resonance,
drops out completely.} Thus, from the recent Mainz experiments on
eta photoproduction off proton and deuteron targets, we can extract,
 in a
model-independent fashion, a ratio of the neutron to proton helicity
amplitudes:
\begin{equation}
A^{n}_{1/2}/A_{1/2}^{p}=\sqrt{\frac{\chi_{p}}{\chi_{n}}}\frac{
\xi_{n}}{\xi_{p}} = -0.84\pm 0.15,
\end{equation}
using $\chi_{p}/\chi_{n} \simeq 0.987$.
 This value compares very favorably
with  recent quark model estimates (e.g. that of
Capstick\cite{ref3}, which yields about $-0.83$).
The  inference of this quantity, done here directly from the
experiments in an essentially model-independent fashion,  is the
{\it central result} of this Letter. The sign of this
quantity is negative, as predicted by the quark models\cite{ref3}.

In Fig.2, we show the differences between proton and neutron targets
in studying polarization observables at the photon lab energy of $780$
MeV, taking an example. We also demonstrate here the effect of
changing the sign of the electromagnetic
helicity amplitude A$_{1/2}$ for neutron to
N$^*$(1535) excitation.
The consequence of the change of
sign of this helicity amplitude for neutron is that
polarization observables, viz., recoil nucleon polarization(RNP),
 polarized
target asymmetry (PTA)
and polarized photon asymmetry(PPA),  {\it all change sign}, with the
PPT showing maximum sensitivity.
Thus, the theoretical
prediction of the sign of this helicity amplitude
 for neutron, of crucial
importance to nuclear excitation of the N$^*$(1535) resonance, can be
verified
by the {\it sign  of the measured polarizations.} The role of the
D13 resonance, N$^*$(1520), is demonstrated in Fig.3.
 It is correlated with
the choice of the $g_{\eta}$. Polarization
experiments for proton and neutron targets
would thus be quite useful. For latter, the polarized $^{3}$He
 target would
serve as a polarized neutron target.

Fig.4 demonstrates the RNP as a function of the photon lab energy
at cm angle
$90^0$, compared with the data of Heusch {\it et al.}\cite{ref11}.
 While the
agreement with proton data is satisfactory, the quality of this data
is not good. More experimental work is needed here.

In summary, we have extracted, using the effective Lagrangian
approach\cite{ref1} and the available new data of eta photoproduction
from proton and deuteron targets\cite{ref2,ref4}, the
 ratio $A_{1/2}^{n}/A_{1/2}^{p}$, both in magnitude
and in sign. In so doing, we have not made the simple
assumption that only
the contribution of N$^*$(1535) be taken into account.
 The extracted ratio,
of fundamental interest in nuclear physics,
agrees nicely with the prediction of recent quark model
estimates; it is  now quite accurate to invite a precise estimate
by lattice QCD methods.
New  polarization experiments, discussed in this Letter,
can provide  independent tests for this ratio and
further theoretical insights into the nucleon resonance excitation.
Our understanding of the eta photoproduction
mechanism on proton and neutron, in turn,
would help that in complex nuclei. The near cancellation of
the sum of the proton and neutron helicity
 amplitudes to excite N$^*$(1535)
would have an important bearing in explaining
the new Mainz experiment,  which does not see
any coherent contribution to the eta photoproduction
 in nuclei\cite{ref10}.
The resultant upper limit for the coherent cross-section,
obtained from this experiment,
supports our conclusion on the sign of
the dominant helicity amplitude, for the N$^*$(1535) excitation, for
neutrons.

One of us (NCM) thanks Dr. Bernd Krusche for his kind hospitality and
valuable discussions. He is also thankful to Drs. C. Bennhold,
W. N\"{o}renberg, C. Sauermann, A. \v{S}varc and
 L. Tiator for helpful comments.
The research at Rensselaer has been supported by the
U. S. Department of Energy; that   at SAL has been supported
by the Natural Sciences and Engineering Research Council of Canada.



\newpage

Fig. 1: {  Angular distributions for eta
photoproduction off protons for two sample values of $E_{\gamma}$,
716 and 775 MeV.
Experimental points (circles) are from \protect{\cite{ref2}}, and the
dot-dashed line is our full effective Lagrangian  fit, while
dots represent resonances alone.
Also shown are our
predictions for the neutron target assuming $A^{n}_{1/2}/A^{p}_{
1/2}=-0.83$ (solid line), vs. the
inferred differential cross-section (stars). Here the long-dashed lines
represent our predictions with resonances alone.}

Fig. 2: {The differential cross-section $d\sigma /d\Omega$,
recoil nucleon polarization (RNP), polarized target asymmetry (PTA)
and polarized photon asymmetry (PPA)  for eta photoproduction off
proton (dot-dashed line) and neutron targets (solid line).
We also display (dashed line) the effect when $A_{1/2}$ changes sign
for the neutron target from the negative value predicted by the
 quark model
for the N$^*$(1535) excitation. The photon lab energy,
chosen here for illustrative purposes, is $780 MeV$.}

Fig. 3: {The comparison between the results of
two different sets of $A_{1/2}$,  $A_{3/2}$
for N$^*$(1520). Solid line is the result
of  fitting our inferred neutron
differential cross-sections from Krusche
{\it et al.}\protect{\cite{ref4}} ($A_{1/2}=43.8$, $A_{3/2}=96.6$ for
N$^*$(1520)). Dashed line is the result using the PDG amplitudes for
N$^*$(1520)($A_{1/2}=-62$, $A_{3/2}=-137$).
Observables are defined in Fig.2,
$E_{\gamma}=780$MeV. All amplitudes here are in the usual
unit\protect{\cite{ref1}}.}

Fig. 4: { Recoil nucleon polarization (dot-dashed line, proton; solid
line, neutron) vs. $E_{\gamma}$, photon lab  energy, at meson cm angle
of $90^0$.  Experimental points for proton are from Heusch
{\it et al}.\protect{\cite{ref11}}.}
\newpage
\begin{table}
\caption{A comparison of various contributions to
  the $E_{0^+}$ multipole, in
units of $10^{-3}/m_{\pi+}$, for the $\gamma +p\rightarrow\eta +p$ and
the $\gamma +n\rightarrow\eta +n$
reactions, at their respective thresholds. The parameters, defined in
\protect\cite{ref1}, $\alpha =+1$ and
 $\Lambda^{2}=1.2 GeV^{2}$ are used.
The targets (Re/Im parts of the $E_{0^+}$ amplitude) are indicated.
 Our model parameters are fitted to the
experiment of Krusche {\it et al.}\protect{\cite{ref2}} for protons,
and our inferred data for neutrons\protect{\cite{ref4}}. A refers to
the estimate by Tiator {\it et al}.\protect{\cite{ref6}}, and B is the
present work.}
\label{table1}
\begin{center}
\begin{tabular}{lcccccccc}
&&&&\\
\multicolumn{1}{l}{Contributions} &\multicolumn{8}{c}
{Targets (Re/Im parts of the $E_{0^+}$ amplitude)}\\
      & p (Re)&p (Re)&p (Im)&p (Im)&n (Re)&n (Re)&n (Im)& n (Im)
 \\
 & A&B&A&B&A&B&A&B\\[1em] \hline
&& \\[1em]
Nucleon Born terms &$-5.2$&$-3.4$ &$0.0$&$0.0$ &$3.6$
  &$2.3$&$0.0$ &$0.0$
\\[1em]
$\rho+\omega$      &$3.0$&$2.9$ &$0.0$  &$0.0$ &$-2.3$
  &$-2.0$&$0.0$ &$0.0$
 \\[1em]
$N^{*}(1535)$      &$12.9$&$12.2$&$6.0$ &$8.5$&$-7.1$
&$-10.5$&$-4.9$&$-7.3$
 \\[1em]
$N^{*}(1520)$      &$--$&$1.6$&$--$  &$0.0$ &$--$
& $-0.2$&$--$  &$0.0$\\[1em]
Total              &$10.7$&$13.3$&$6.0$ &$8.5$&$-5.9$
& $-10.4$&$-4.9$
 &$-7.3$ \\[1em]
\end{tabular}
\end{center}
\end{table}

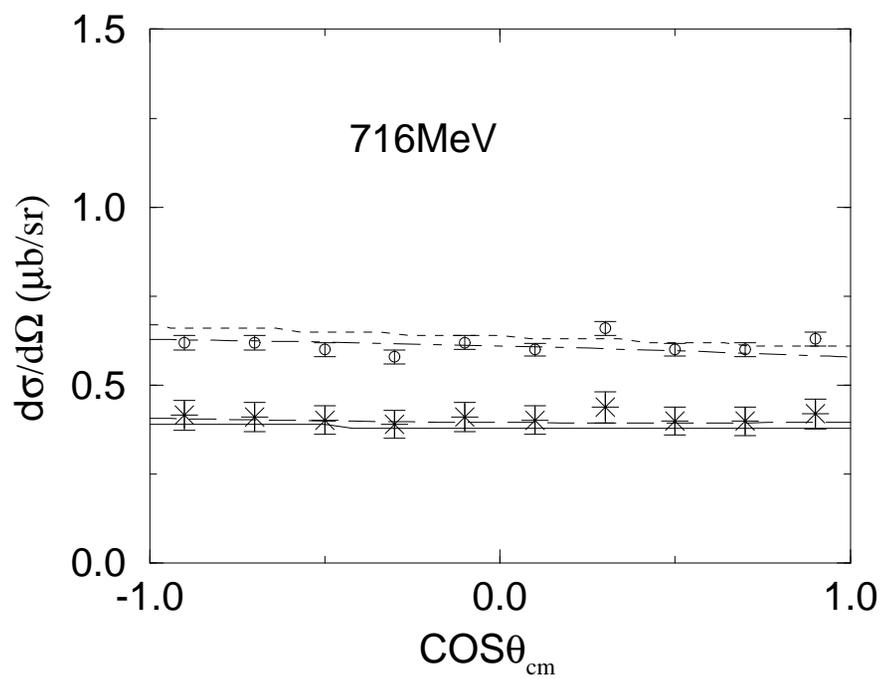
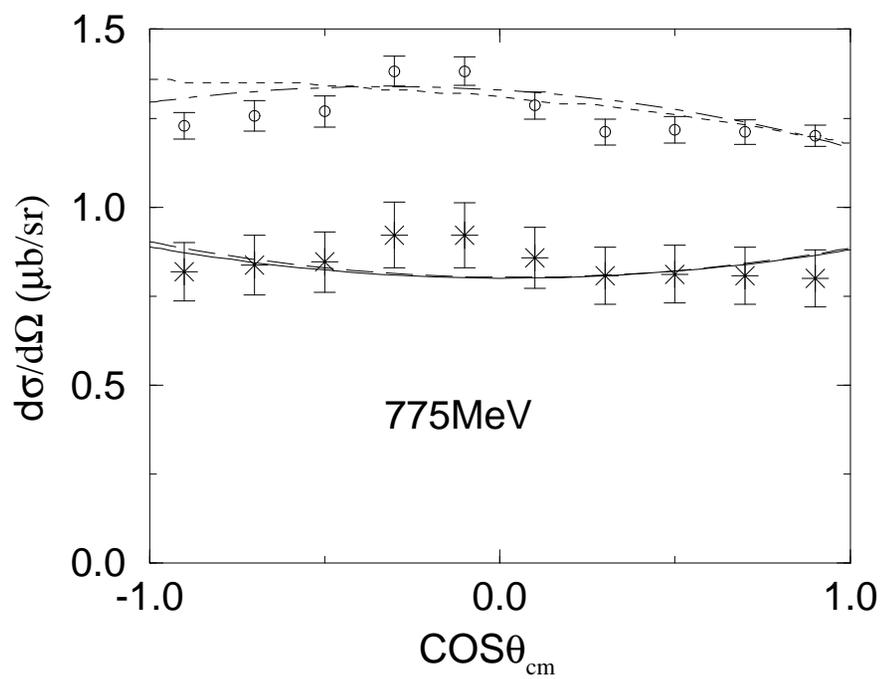
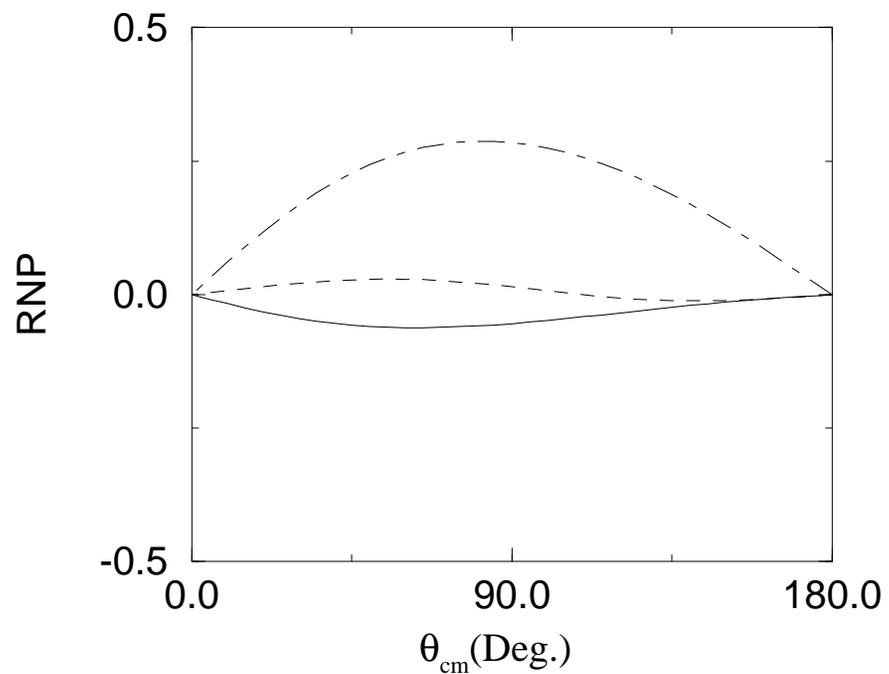
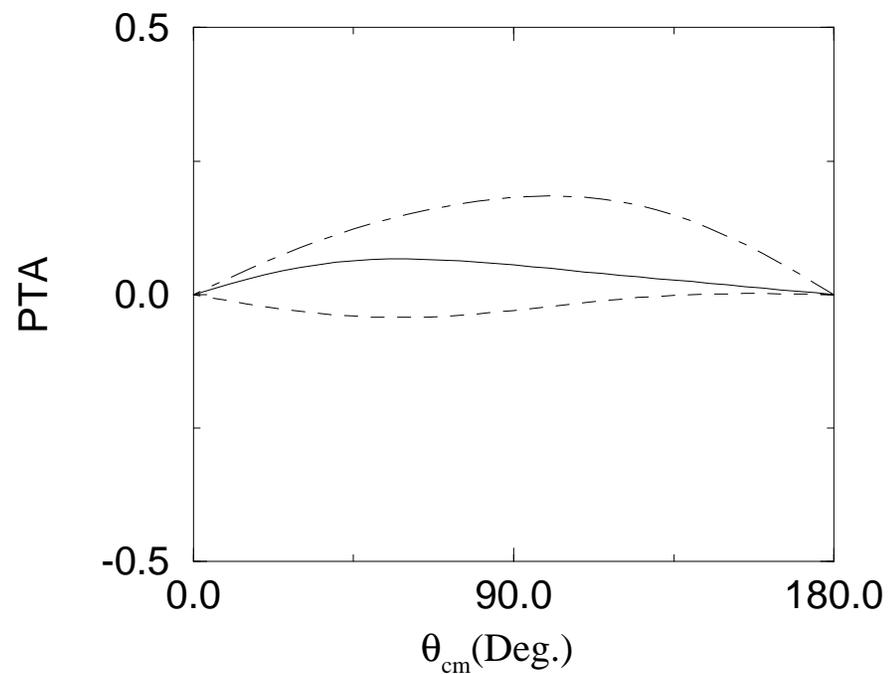
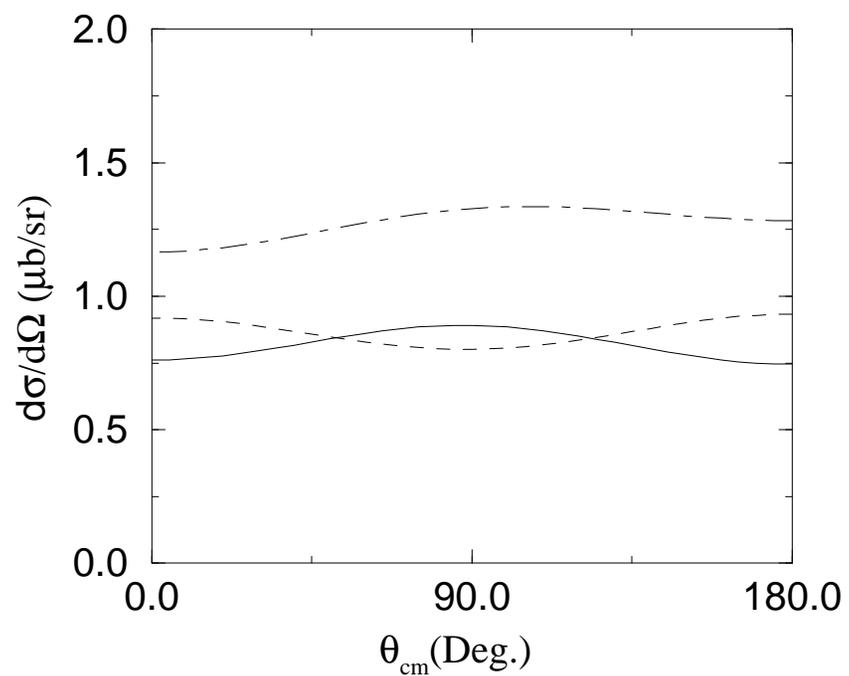
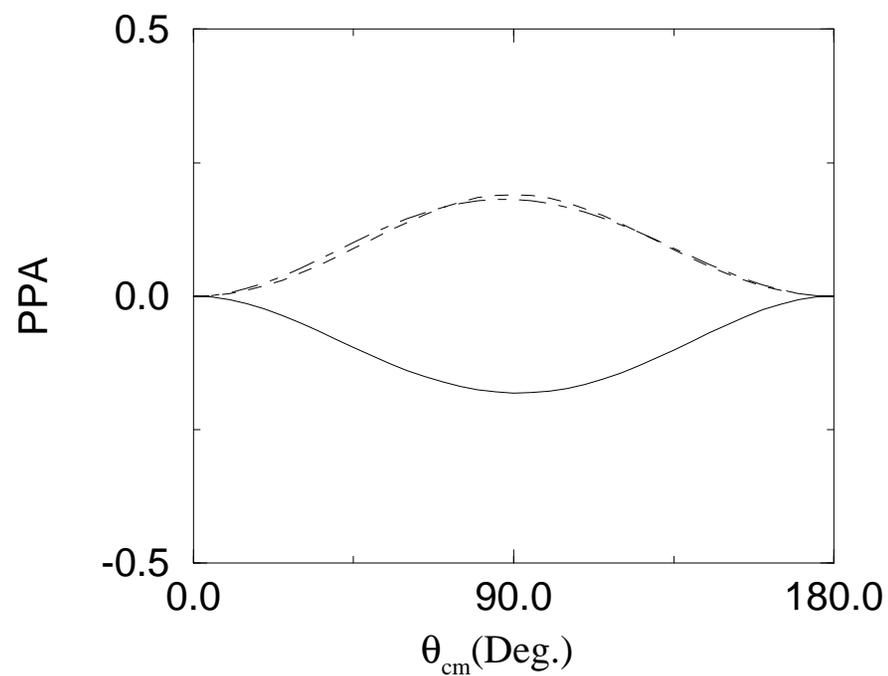
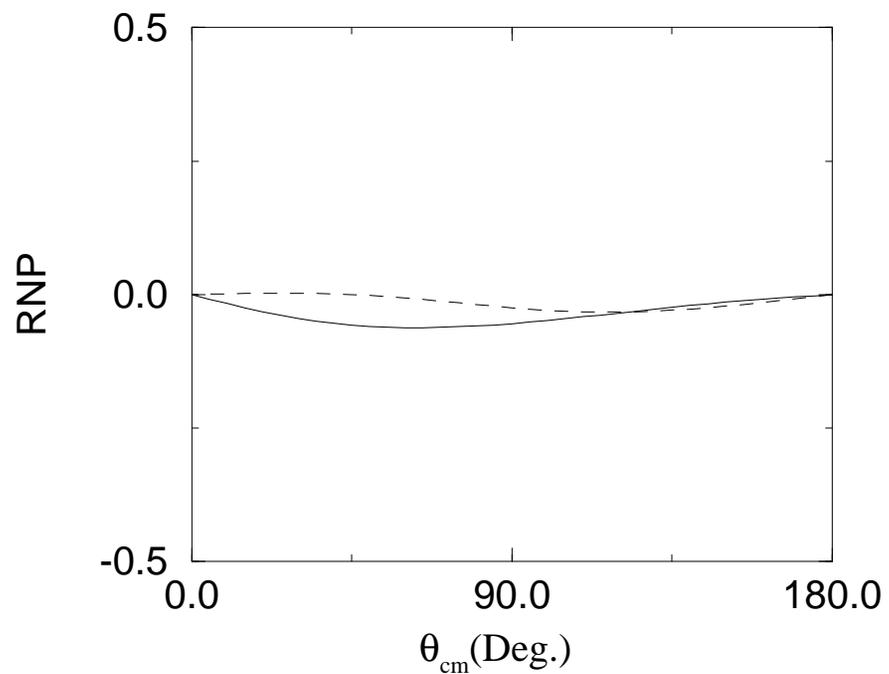
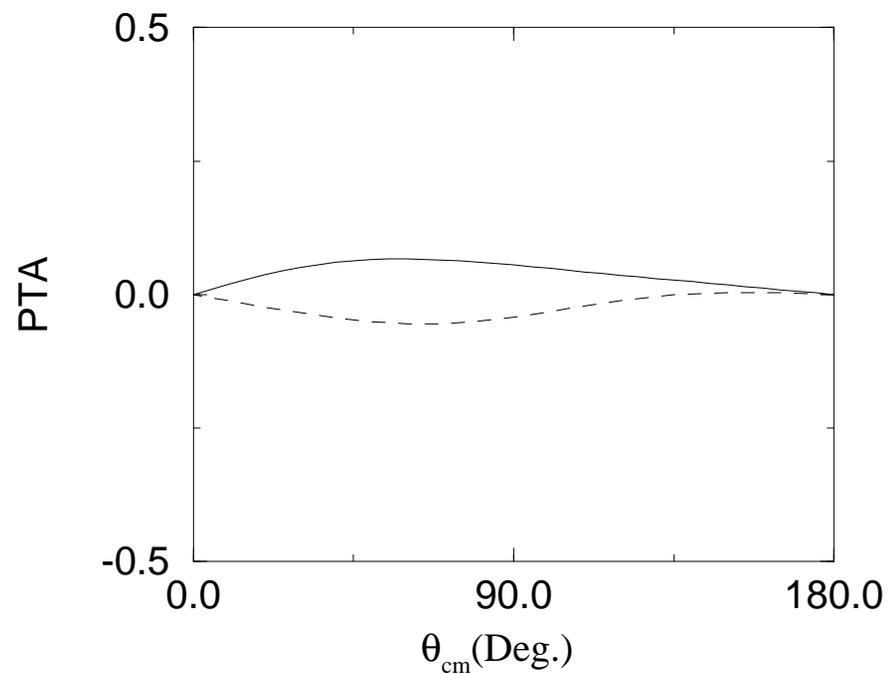
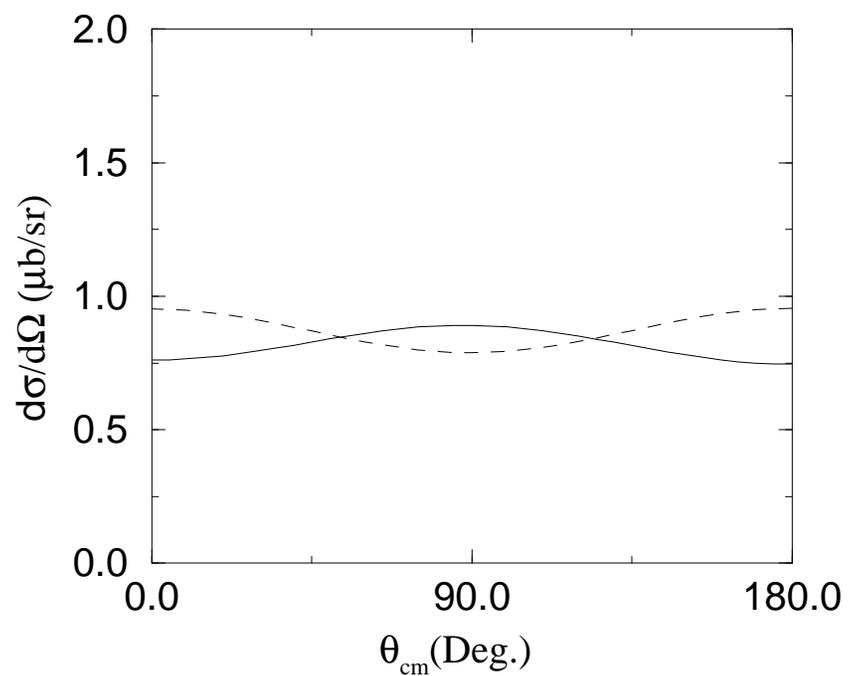
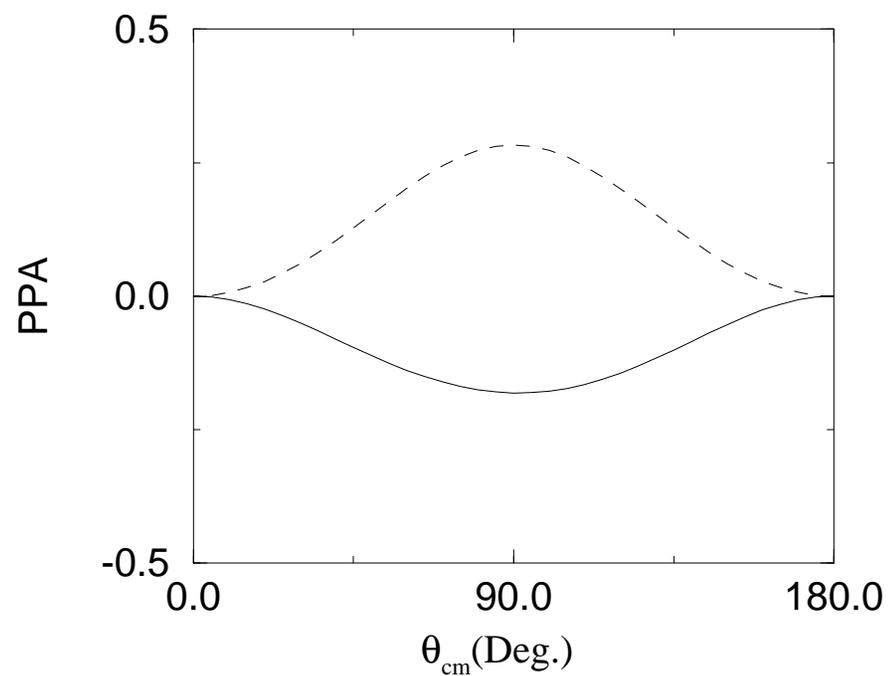
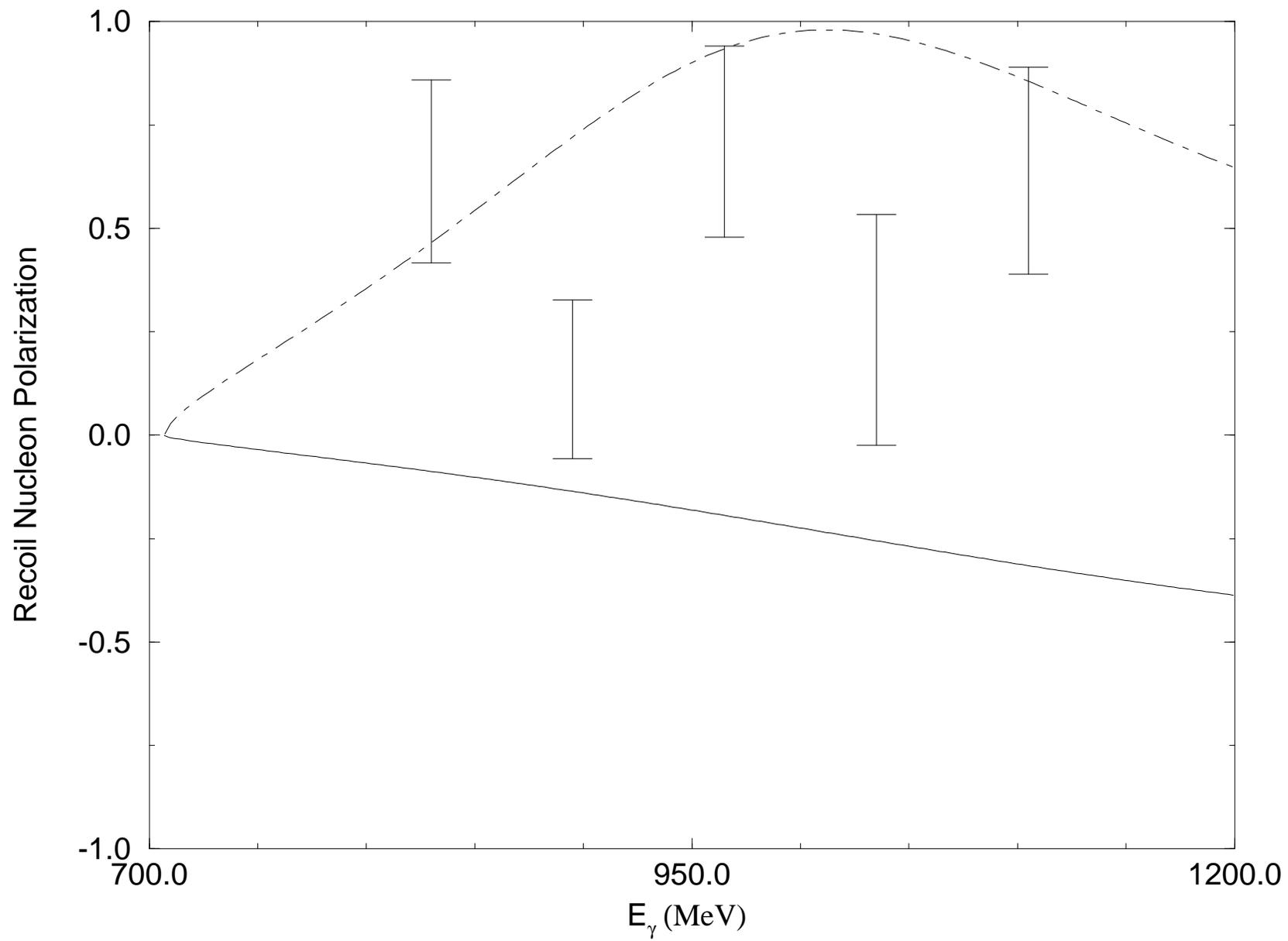
\end{document}